\documentclass{article}
\usepackage{spconf,amsmath,graphicx,hyperref}
\usepackage{multirow}
\usepackage{comment}
\usepackage{makecell}
\usepackage{booktabs}
\usepackage{orcidlink}

\title{Robust Cross-Domain Speech-Based Alzheimer's Disease Detection via Iterative Adversarial Self-Training}
\name{\shortstack{
Luqi Sun\orcidlink{0009-0004-4702-0508}$^1$,
Shreeram Suresh Chandra\orcidlink{0000-0002-1085-2544}$^1$,
Aurosweta Mahapatra\orcidlink{0009-0001-3057-2021}$^1$,\\
Emily Mower Provost\orcidlink{0000-0003-1870-6063}$^2$,
Brian MacWhinney\orcidlink{0000-0002-4988-1342}$^3$,
Berrak Sisman\orcidlink{0000-0001-8078-3305}$^1$
}}

\address{
$^1$ Center for Language and Speech Processing (CLSP), Johns Hopkins University, USA\\
$^2$ University of Michigan, Ann Arbor, USA\space\space\space\space
$^3$ Carnegie Mellon University, USA
}

\begin{document}

\maketitle

\begin{abstract}
As Alzheimer’s disease (AD) has increasingly become a major global public health issue, speech-based AD detection has attracted widespread attention. However, most existing methods are trained and evaluated on a single dataset, often leading to severe cross-domain performance degradation due to reliance on dataset-specific artifacts rather than disease-related speech cues. In real-world applications, reliable Alzheimer’s disease detection requires models that are robust to variations in recording environments, speakers and data collection conditions. To address this challenge, this paper adopts unsupervised domain adaptation to learn robust, domain-invariant feature representations in the absence of target-domain diagnosis labels. On this basis, a novel unsupervised domain adaptation method, Iterative Adversarial Self-Training (IAST), is proposed. Results demonstrate that IAST significantly improves the generalization ability and robustness under various cross-domain settings.\footnote{We release our codes and models with this submission: https://website5544.github.io/website/} 
\end{abstract}
\begin{keywords}
Alzheimer’s disease detection, Speech
\end{keywords}
%
\section{Introduction}
\label{sec:intro}

In recent years, deep learning has become the dominant approach for speech-based AD detection. Prior work has explored a wide range of architectures and representations, including end-to-end automatic speech recognition frameworks~\cite{LinICASSP2024}, transformer-based models for speech-level AD recognition~\cite{DongHAFFormerICASSP2024}, and hybrid approaches that incorporate disfluency-related paralinguistic features~\cite{JinConsenICASSP2023}. In parallel, a series of benchmark challenges organized at Interspeech and ICASSP, such as ADReSS~\cite{ADReSS}, ADReSSo~\cite{ADReSSo}, ADReSS-M~\cite{LuzMultilingualICASSP2023}, have further accelerated progress in this area. Collectively, these studies demonstrate that speech-based AD detection models can achieve strong performance when trained and evaluated under matched, single-dataset conditions.

However, in real-world applications, AD detection often needs to cope with different recording environments, device conditions, and data collection processes. Although existing models can achieve high detection performance when trained and evaluated on a single dataset, their performance often degrades substantially when applied to data collected under different conditions, revealing limited robustness. This degradation arises in part because models trained on a single dataset may learn to rely on dataset-specific recording characteristics rather than disease-related speech features. Prior work has provided clear evidence of this behavior: Liu et al.~\cite{LiuCleverHans2024} demonstrated that models trained using only silent segments from the Pitt Corpus~\cite{BeckerArchNeurol1994} can still achieve extremely high detection accuracy on the same dataset. This suggests the presence of a Clever Hans effect, in which models exploit recording recording-environment artifacts instead of clinically meaningful speech cues. 

Our cross-domain experiments further confirm these limitations: across multiple model architectures, we observe consistent drops in recognition accuracy under cross-domain evaluation, with a maximum decrease of 34.38\% (see Table~\ref{tabs:results}). These results indicate that AD detection models trained and evaluated solely on a single dataset exhibit limited cross-domain generalization, significantly constraining their practical applicability. To address this challenge, this paper investigates unsupervised domain adaptation for speech-based AD detection. 

Recent advances in self-supervised speech representation learning~\cite{bert} have shown that large pre-trained models can capture robust and transferable speech features across speakers and recording conditions. In particular, XLSR~\cite{ConneauXLSR2020}, pre-trained on large-scale multilingual unlabeled speech data, has demonstrated strong performance in AD detection and related speech-based cognitive impairment tasks~\cite{xlsr1}. These properties make self-supervised speech representations especially promising for cross-domain Alzheimer’s disease detection, where recording conditions and data distributions often vary substantially. Motivated by these insights, this work investigates how different levels of speech representation and unsupervised domain adaptation strategies can be combined to improve robustness and generalization in cross-domain settings. 




\vspace{-3mm}
\section{Dataset}
\vspace{-2mm}

We use three public datasets: the Pitt Corpus~\cite{BeckerArchNeurol1994}, the ADReSS dataset~\cite{ADReSS}, and the English Lu Corpus, all accessed through DementiaBank~\cite{DementiaBank}. We select the Cookie Theft picture description task because it elicits spontaneous speech, in which paralinguistic markers provide strong cues for Alzheimer’s disease detection.

\textbf{Pitt Corpus}: The Pitt Corpus contains 306 speech samples from individuals with Alzheimer’s disease and 243 from healthy controls. 

\textbf{ADReSS dataset}: Released as part of the ADReSS Challenge~\cite{ADReSS}, this dataset includes 78 speech samples from individuals with Alzheimer’s disease and 78 from healthy controls, balanced for age and gender. All recordings have undergone noise reduction and volume normalization. 

\textbf{English Lu Corpus}: Recordings longer than 60 seconds are segmented into 60-second clips, while segments shorter than 15 seconds are discarded. After preprocessing, 74 speech samples are obtained, including 38 samples from individuals with Alzheimer's disease and 36 from healthy controls.

\section{Methodology}
\label{sec:method}

This section presents the designed AD detection models, traditional unsupervised domain adaptation methods, and the proposed Iterative Adversarial Self-Training (IAST).

\vspace{-4mm}
\subsection{Models}
\label{sec:Models}

We implement three speech-based Alzheimer’s disease detection models with increasing representational capacity: an eGeMAPS-based model using handcrafted acoustic features, a pre-trained embedding-based model with frozen parameters, and a fine-tuned pre-trained model for the target task. 


\textbf{eGeMAPS-based model}: 
Each input utterance is first segmented into 10 equal segments, from which 25-dimensional eGeMAPS~\cite{EybenGeMAPS2015} acoustic features are extracted using the OpenSMILE toolkit. Two linear layers are then applied to remap the feature dimensions, followed by an attention pooling layer to aggregate segment-level representations. Finally, a linear classification layer outputs the AD prediction. 

\textbf{Frozen-parameter XLSR model}: 
The audio input is fed into a pre-trained XLSR model~\cite{ConneauXLSR2020} with frozen parameters to extract frame-level speech embeddings. Five linear layers are then used to project the embeddings, which are aggregated using an attention pooling layer for temporal feature representation. A padding mask~\cite{VaswaniTransformer2017} is applied to ensure that padded time steps do not contribute to the pooled representation. Finally, a linear classification layer outputs the AD prediction.

\textbf{Fine-tuned XLSR model}: 
The XLSR model is fine-tuned separately on the Pitt Corpus and the ADReSS dataset. To mitigate catastrophic forgetting and reduce the risk of overfitting, only the final three Transformer layers are updated during fine-tuning, while the remaining layers are kept frozen. The classification head is identical to that of the Frozen-parameter XLSR model.

  

\vspace{-1mm}
\subsection{Domain Adversarial Training}

Domain Adversarial Training (DAT)~\cite{GaninDANN2016} is used to learn domain-invariant representations from the labeled source and unlabeled target domains. A domain discriminator is jointly trained with the AD classifier through adversarial learning. While the classifier minimizes the source-domain classification loss, the feature extractor learns to confuse the domain discriminator, thereby aligning the source and target feature distributions without target labels. DAT is applied to all three models described in Section~\ref{sec:Models}.

\vspace{-1mm}
\subsection{Self Training}

Self-Training (ST)~\cite{ScudderIT1965} is employed to leverage unlabeled target-domain data through a pseudo-labeling mechanism. This training strategy is also applied consistently to all three models. Models are first trained using labeled source-domain data and then used to generate predictions for target-domain samples. High-confidence predictions are selected as pseudo-labels and combined with the source-domain data for subsequent training.  This process is repeated in an iterative manner, enabling the model to progressively incorporate more reliable target-domain supervisory information, thereby continuously improving its performance on the target domain and enhancing cross-domain generalization ability.

\begin{figure}[t]
  \centering
  \includegraphics[width=\linewidth]{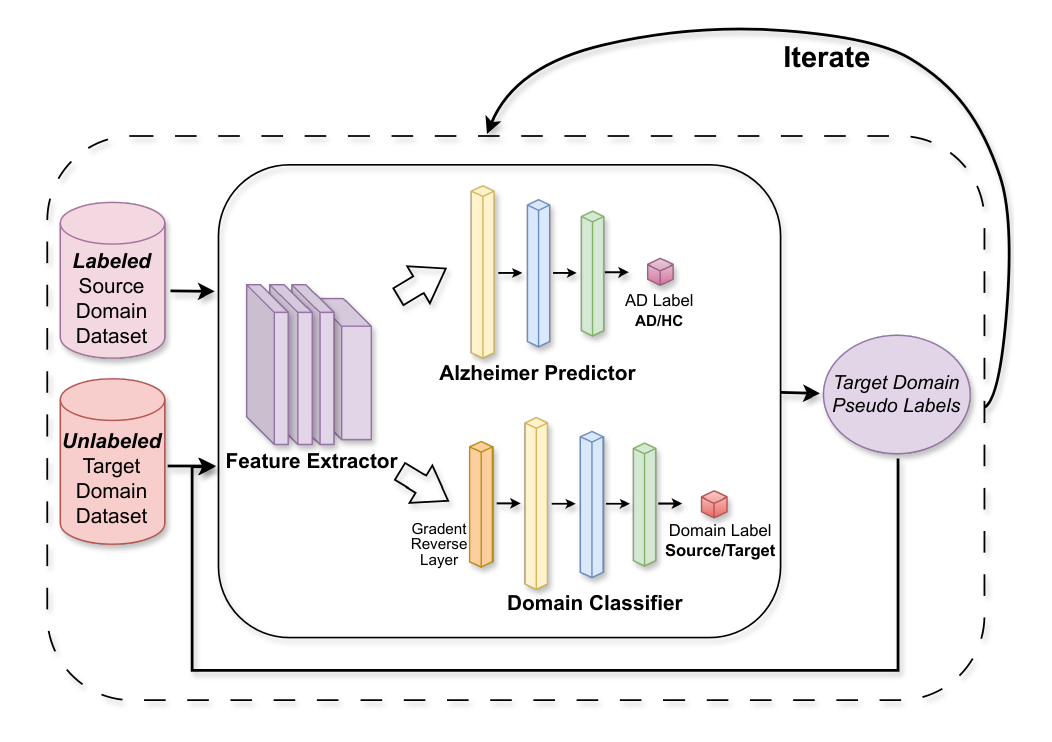}
  \vspace{-7mm}
  \caption{Iterative Adversarial Self-Training pipeline (AD: Alzheimer’s disease; HC: healthy controls).}
  \label{IAST Pipeline}
  \vspace{-4mm}
\end{figure}

\vspace{-1mm}
\subsection{Iterative Adversarial Self-Training}

Iterative Adversarial Self-Training. IAST alternates DAT and ST to jointly improve feature alignment and target-domain pseudo-label quality, as shown in Fig.~\ref{IAST Pipeline}. DAT first aligns labeled source and unlabeled target representations. The aligned model then generates high-confidence target pseudo-labels for self-training. The updated model is subsequently used to restart adversarial alignment and pseudo-label generation. Repeating this cycle progressively incorporates reliable target samples while maintaining domain-invariant representations.

\vspace{-1mm}
\section{Experiments}
As described in Section~\ref{sec:method}, we evaluate speech-based Alzheimer’s disease detection models under both baseline and unsupervised domain adaptation settings. In the baseline setting, models are trained on a single source dataset—either the Pitt Corpus~\cite{BeckerArchNeurol1994} or the ADReSS dataset~\cite{ADReSS}—and directly evaluated on the English Lu Corpus to assess cross-domain generalization without adaptation. In the UDA setting, the Pitt~\cite{BeckerArchNeurol1994} and ADReSS~\cite{ADReSS} datasets are used as source domains, while the Lu corpus serves as the target domain. No target-domain labels are used. Models are trained using Domain Adversarial Training, Self-Training, or the proposed Iterative Adversarial Self-Training. For the source domain, 80\% of labeled data is used for training and 20\% for validation. For the target domain, 80\% of unlabeled data is used for domain alignment or pseudo-label generation, and the remaining 20\% is reserved for evaluation. 


\vspace{-1mm}
\section{Results and Analysis}

\begin{table*}[t]
\centering
\caption{Cross-domain performance comparison of speech-based Alzheimer’s disease detection models under different unsupervised domain adaptation methods (DAT: Domain Adversarial Training; ST: Self-Training; IAST: Iterative Adversarial Self-Training).}
\label{tabs:results}
\small
\begin{tabular}{p{3cm}lllcccc}
\toprule

\multirow{2}{*}{\textbf{Model}} 
& \multirow{2}{*}{\textbf{Source Domain}} 
& \multirow{2}{*}{\textbf{Target Domain}} 
& \multirow{2}{*}{\textbf{Domain Adaptation}} 
& \multicolumn{2}{c}{\textbf{Within Domain}} 
& \multicolumn{2}{c}{\textbf{Cross Domain}} \\

\cmidrule(lr){5-8}

 &  &  &  
& \textbf{Src Acc} & \textbf{Src F1} 
& \textbf{Tgt Acc} & \textbf{Tgt F1} \\

\midrule
\multirow{8}{*}{eGeMAPS-based model}  & \multirow{4}{*}{ADReSS} & \multirow{4}{*}{Lu} & -- & 65.62\% & 0.6667 & 54.05\% & 0.4516 \\
         &  &  & DAT & 62.50\% & 0.6667 & 62.50\% & 0.7000 \\
          &  &  & ST & 68.75\% & 0.7222 & 62.50\% & 0.6667 \\
         &  &  & \textbf{IAST \textit{(proposed)}} & \textbf{71.88\%}& \textbf{0.7429} & \textbf{75.00\%} & \textbf{0.7778} \\
\cmidrule(lr){2-8}
 & \multirow{4}{*}{Pitt}   & \multirow{4}{*}{Lu} & -- & 61.26\% & \textbf{0.7190} & 58.11\% & 0.6931 \\
        &    &  & DAT & \textbf{62.16\%} & 0.7042 & 62.50\% & 0.7273 \\
        &    &  & ST  & 59.46\% & 0.7134 & 62.50\% & 0.7273 \\
        &    &  & \textbf{IAST \textit{(proposed)}} & 58.56\% & 0.6462 & \textbf{68.75\%} & \textbf{0.7368} \\
\midrule
\multirow{8}{*}{\makecell{XLSR-based model \\ \textit{(Frozen Parameters)}}} & \multirow{4}{*}{ADReSS} & \multirow{4}{*}{Lu} & -- & \textbf{87.50\%} & \textbf{0.8750} & 68.92\% & 0.6102 \\
            &    &    & DAT & 81.25\% & 0.8421 & 81.25\% & 0.8235 \\
            &    &    & ST  & 84.38\% & 0.8571 & 75.00\% & 0.7500 \\
            &    &    & \textbf{IAST \textit{(proposed)}} & 81.25\% & 0.8235 & \textbf{81.25\%} & \textbf{0.8235} \\
\cmidrule(lr){2-8}
  & \multirow{4}{*}{Pitt}   & \multirow{4}{*}{Lu} & -- & 70.72\% & 0.7442 & 58.11\% & 0.6804 \\
            &      &    & DAT & 70.27\% & 0.7227 & 75.00\% & 0.7778 \\
            &      &    & ST  & \textbf{71.17\%} & \textbf{0.7576} & 81.25\% & 0.8421 \\
            &      &    & \textbf{IAST \textit{(proposed)}} & 67.57\% & 0.7568 & \textbf{87.50\% }& \textbf{0.8750} \\
\midrule
\multirow{4}{*}{\makecell{XLSR-based model \\ \textit{(Fine-tuned on} \\ \textit{ADReSS Dataset)} \\ }} & \multirow{4}{*}{ADReSS} & \multirow{4}{*}{Lu}  & -- & 84.38\% & 0.8485 & 50.00\% & 0.4789 \\
                      &    &    & DAT & 81.25\% & 0.8235 & 68.75\% & 0.6154 \\
                      &    &    & ST  & 84.38\% & 0.8387 & 81.25\% & 0.8235 \\
                      &    &    & \textbf{IAST \textit{(proposed)}}  & \textbf{87.50\%} & \textbf{0.8667} & \textbf{81.25\%} & \textbf{0.8235} \\
\midrule
\multirow{4}{*}{\makecell{XLSR-based model \\ \textit{(Fine-tuned on} \\ \textit{Pitt Corpus)} \\ }}& \multirow{4}{*}{Pitt} & \multirow{4}{*}{Lu} & -- & 69.37\% & 0.7213 & 63.51\% & 0.6494 \\
                    &    &    & DAT & 67.57\% & 0.7313 & 75.00\% & 0.7500 \\
                    &    &    & ST  & 70.27\% & 0.7317 & 87.50\% & 0.8571 \\
                    &    &    & \textbf{IAST \textit{(proposed)}} & \textbf{70.27\%} & \textbf{0.7660} & \textbf{87.50\%} & \textbf{0.8571} \\
\bottomrule
\end{tabular}
\vspace{-5mm}
\end{table*}

\textbf{Baseline Cross-Domain Performance.} Table~\ref{tabs:results} reports cross-domain results under the baseline setting, where models are trained on a single source dataset and evaluated on the target domain without adaptation. A consistent performance gap is observed across all architectures. For instance, the ADReSS-finetuned XLSR model achieves 84.38\% accuracy in-domain but drops to 50.00\% on the Lu corpus, which is a 34.38\% decrease. Similarly, the Frozen-XLSR model trained on Pitt decreases from 70.72\% to 58.11\%, and the eGeMAPS-based model drops from 65.62\% to 54.05\% when transferred from ADReSS to Lu. These results show that strong in-domain performance does not guarantee cross-domain generalization. Even large pretrained models such as XLSR suffer substantial degradation under domain shift, suggesting reliance on source-domain–specific characteristics rather than stable disease-related speech features.

\textbf{Traditional Domain Adaptation Methods.} We next examine the effect of traditional unsupervised domain adaptation methods on cross-domain AD detection performance. As shown in Table~\ref{tabs:results}, both Domain Adversarial Training and Self-Training improve target-domain performance across most model architectures compared to the baseline setting, although the degree of improvement varies substantially by model. For the eGeMAPS-based model trained on the ADReSS dataset, DAT improves target-domain accuracy from 54.05\% to 62.50\%, while ST yields a similar improvement to 62.50\%. When trained on the Pitt Corpus, both DAT and ST again improve target-domain accuracy relative to the baseline, indicating that feature alignment and pseudo-labeling can partially help domain mismatch even for handcrafted acoustic features. More pronounced gains are observed for models based on pre-trained speech representations. For example, the XLSR-based model with frozen parameters trained on the Pitt Corpus improves from 58.11\% accuracy on the target domain under the baseline setting to 75.00\% with DAT and 81.25\% with ST. Similarly, for the XLSR-based model fine-tuned on the ADReSS dataset, ST increases target-domain accuracy from 50.00\% to 81.25\%. These results suggest that self-training is particularly effective when the underlying model produces sufficiently reliable target-domain predictions, as is often the case for pre-trained representation-based models.


\textbf{The proposed IAST.} We next analyze the effect of the proposed IAST method on cross-domain AD detection.  As shown in Table~\ref{tabs:results}, IAST consistently yields the strongest target-domain performance across different model architectures and source–target settings, outperforming or matching the best results achieved by traditional unsupervised domain adaptation methods. For the eGeMAPS-based model trained on the ADReSS dataset, IAST improves target-domain accuracy to 75.00\%, substantially exceeding the performance of both DAT and ST (62.50\%). When trained on the Pitt Corpus, IAST again achieves the highest target-domain accuracy of 68.75\%, compared to 62.50\% obtained with DAT and ST. These results indicate that the proposed iterative combination of adversarial feature alignment and self-training is particularly beneficial for models with limited representational capacity. For models based on representations extracted from large-scale pre-trained model, IAST also demonstrates clear advantages. In the case of the XLSR-based model with frozen parameters trained on the Pitt Corpus, IAST increases target-domain accuracy to 87.50\%, outperforming both DAT (75.00\%) and ST (81.25\%). Similarly, for the XLSR-based model fine-tuned on the ADReSS, IAST improves target-domain accuracy to 81.25\%, exceeding the result obtained with DAT and matching the best performance achieved by ST. 

\begin{figure}[t]
  \centering
  \includegraphics[width=0.7\linewidth]{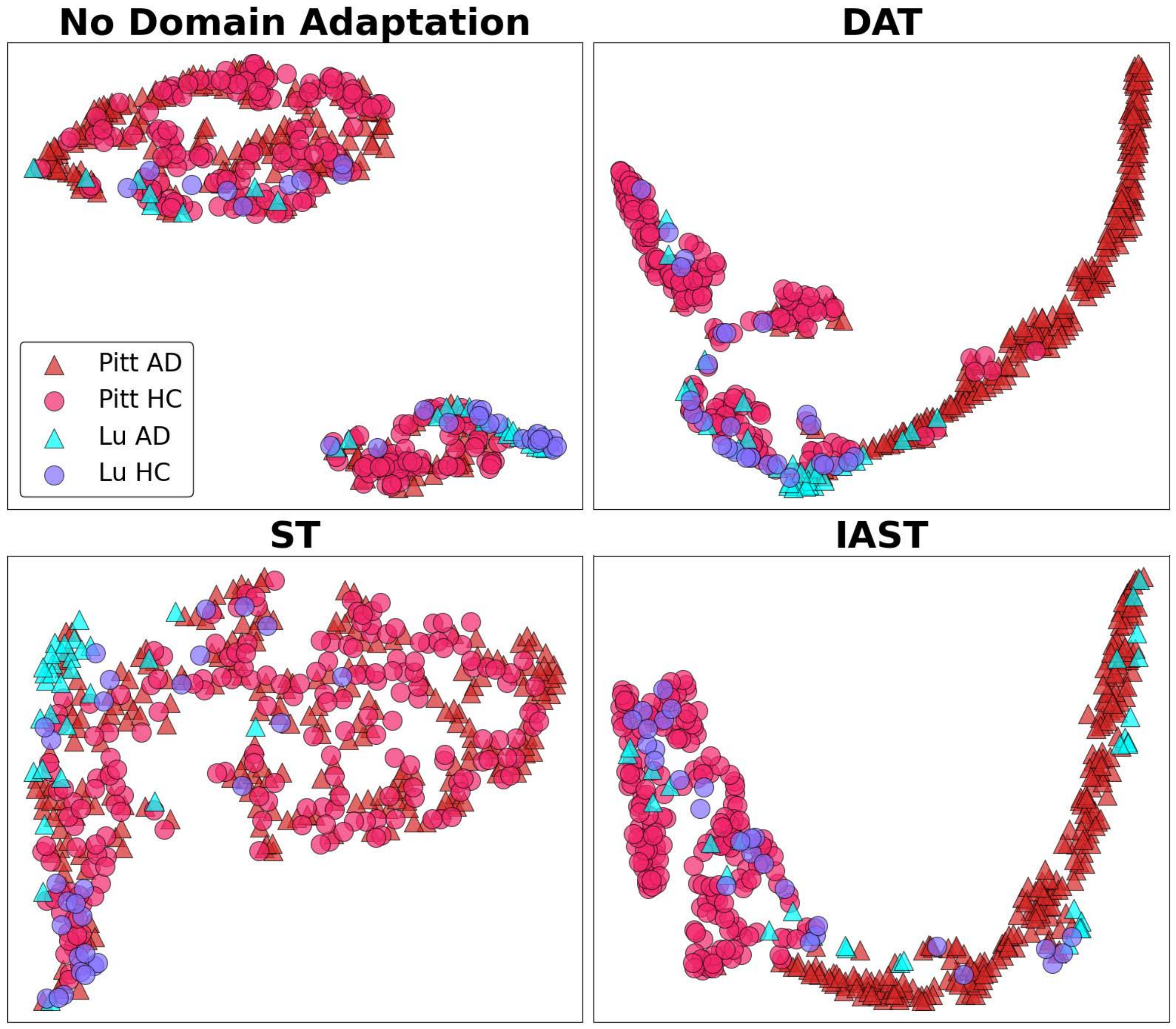}
  \caption{t-SNE visualization of embeddings from the XLSR-based model fine-tuned on Pitt Corpus.}
  \label{embedding}
  \vspace{-5mm}
\end{figure}

\textbf{Embedding Space.} Figure~\ref{embedding} visualizes the embedding distribution of the Pitt-finetuned XLSR model using t-SNE. To quantitatively assess clustering behavior, Table~\ref{tab:cluster} reports two metrics, \textbf{\textit{silhouette score}}~\cite{Silhouette} and \textbf{\textit{separation ratio}}~\cite{seperation}, under different domain adaptation methods. We evaluate two clustering structures: (i) domain separation (Pitt Corpus vs. English Lu Corpus) and (ii) class separation (Alzheimer’s disease vs. Healthy Controls). The results show that IAST substantially reduces inter-domain discrepancy while simultaneously increasing class separability. This indicates that IAST not only aligns feature distributions across datasets but also enhances discriminative structure in the embedding space.

\begin{table}[htbp]
\centering
\caption{Clustering evaluation from the XLSR-based model fine-tuned on Pitt Corpus (AD: Alzheimer's disease; HC: healthy controls; SS: Silhouette score; SR: Separation ratio).}

\small
\begin{tabular}{p{2.5cm}lcccc}
\hline
\multirow{2}{*}{\textbf{Domain adaptation}}
 & \multicolumn{2}{c}{\textbf{Pitt vs Lu}} & \multicolumn{2}{c}{\textbf{AD vs HC}} \\
\cline{2-3} \cline{4-5}
& \textbf{SS $\downarrow$} & \textbf{SR $\downarrow$}
& \textbf{SS $\uparrow$} & \textbf{SR $\uparrow$} \\
\hline
--  & 0.2062 & 0.9612 & 0.0189 & 0.2318 \\
DAT   & \textbf{0.0010} & 0.9791 & 0.3707 & 1.4624 \\
ST  & 0.1218 & 0.9747 & 0.0069 & 0.1299 \\
\textbf{IAST \textit{(proposed)}} & 0.0031 & \textbf{0.0892} & \textbf{0.4857} & \textbf{1.9484} \\
\hline
\end{tabular}
\label{tab:cluster}
\vspace{-3mm}
\end{table}

\textbf{Source–Target Performance Trade-offs.} While unsupervised domain adaptation improves target-domain performance, it can slightly reduce source-domain accuracy. Compared to the baseline, DAT, ST, and IAST occasionally exhibit modest source-domain degradation. For example, for the frozen XLSR model trained on the Pitt Corpus, source-domain accuracy decreases from 70.72\% to 67.57\% with IAST, while target-domain accuracy increases substantially from 58.11\% to 87.50\%. This trend reflects a trade-off in which improved cross-domain generalization comes at the cost of reduced specialization to source-domain characteristics, favoring more domain-invariant and robust representations.

\vspace{-5mm}
\section{Conclusion}
\vspace{-1mm}

This study demonstrates that speech-based AD detection models trained on a single dataset often exhibit limited robustness when evaluated across datasets, which limits their practical applicability. Unsupervised domain adaptation mitigates this limitation, while the proposed Iterative Adversarial Self-Training (IAST) improves generalization by iteratively combining domain adversarial training with self-training.

\bibliographystyle{IEEEbib}
\bibliography{strings,refs}

@inproceedings{LinICASSP2024,
  title={{Dementia Assessment Using Mandarin Speech with an Attention-Based Speech Recognition Encoder}},
  author={Lin, Zih-Jyun and Chen, Yi-Ju and Kuo, Po-Chih and Huang, Likai and Hu, Chaur-Jong and Chen, Cheng-Yu},
  booktitle={International Conference on Acoustics, Speech and Signal Processing (ICASSP)},
  pages={12461--12465},
  year={2024},
  organization={IEEE}
}

@inproceedings{DongHAFFormerICASSP2024,
  title={{HAFFormer: A hierarchical attention-free framework for Alzheimer’s disease detection from spontaneous speech}},
  author={Dong, Zhongren and Zhang, Zixing and Xu, Weixiang and Han, Jing and Ou, Jianjun and Schuller, Bj{\"o}rn W},
  booktitle={International Conference on Acoustics, Speech and Signal Processing (ICASSP)},
  pages={11246--11250},
  year={2024},
  organization={IEEE}
}

@inproceedings{JinConsenICASSP2023,
  title={{Consen: Complementary and simultaneous ensemble for alzheimer’s disease detection and mmse score prediction}},
  author={Jin, Longbin and Oh, Yealim and Kim, Hyunseo and Jung, Hyuntaek and Jon, Hyo Jin and Shin, Jung Eun and Kim, Eun Yi},
  booktitle={International Conference on Acoustics, Speech and Signal Processing (ICASSP)},
  pages={1--2},
  year={2023},
  organization={IEEE}
}

@inproceedings{ADReSSo,
title = {Detecting Cognitive Decline Using Speech Only: The ADReSSo Challenge},
author = {Saturnino Luz and Fasih Haider and {Sofia de la Fuente} and Davida Fromm and Brian MacWhinney},
year = {2021},
booktitle = {Interspeech},
pages = {3780--3784}
}

@article{BeckerArchNeurol1994,
  title={{The natural history of Alzheimer's disease: description of study cohort and accuracy of diagnosis}},
  author={Becker, James T and Boiler, Fran{\c{c}}ois and Lopez, Oscar L and Saxton, Judith and McGonigle, Karen L},
  journal={Archives of neurology},
  volume={51},
  number={6},
  pages={585--594},
  year={1994},
  publisher={American Medical Association}
}

@inproceedings{LiuCleverHans2024,
  title     = {{Clever Hans Effect Found in Automatic Detection of Alzheimer's Disease through Speech}},
  author    = {Yin-Long Liu and Rui Feng and Jia-Hong Yuan and Zhen-Hua Ling},
  year      = {2024},
  booktitle = {{Interspeech}},
  pages     = {2435--2439}
}

@article{EybenGeMAPS2015,
  title={{The Geneva minimalistic acoustic parameter set (GeMAPS) for voice research and affective computing}},
  author={Eyben, Florian and Scherer, Klaus R and Schuller, Bj{\"o}rn W and Sundberg, Johan and Andr{\'e}, Elisabeth and Busso, Carlos and Devillers, Laurence Y and Epps, Julien and Laukka, Petri and Narayanan, Shrikanth S and others},
  journal={IEEE transactions on affective computing},
  volume={7},
  number={2},
  pages={190--202},
  year={2015},
  publisher={IEEE}
}

@inproceedings{ConneauXLSR2020,
  title     = {{Unsupervised Cross-Lingual Representation Learning for Speech Recognition}},
  author    = {Alexis Conneau and Alexei Baevski and Ronan Collobert and Abdelrahman Mohamed and Michael Auli},
  year      = {2021},
  booktitle = {Interspeech},
  pages     = {2426--2430}
}

@article{DementiaBank,
  title={{DementiaBank: Theoretical rationale, protocol, and illustrative analyses}},
  author={Lanzi, Alyssa M and Saylor, Anna K and Fromm, Davida and Liu, Houjun and MacWhinney, Brian and Cohen, Matthew L},
  journal={American Journal of Speech-Language Pathology},
  volume={32},
  number={2},
  pages={426--438},
  year={2023},
  publisher={American Speech-Language-Hearing Association}
}

@article{GaninDANN2016,
  title={{Domain-adversarial training of neural networks}},
  author={Ganin, Yaroslav and Ustinova, Evgeniya and Ajakan, Hana and Germain, Pascal and Larochelle, Hugo and Laviolette, Fran{\c{c}}ois and March, Mario and Lempitsky, Victor},
  journal={Journal of machine learning research},
  volume={17},
  number={59},
  pages={1--35},
  year={2016},
  publisher = {ACM}
}

@article{VaswaniTransformer2017,
  title={{Attention is all you need}},
  author={Vaswani, Ashish and Shazeer, Noam and Parmar, Niki and Uszkoreit, Jakob and Jones, Llion and Gomez, Aidan N and Kaiser, {\L}ukasz and Polosukhin, Illia},
  journal={Advances in neural information processing systems},
  volume={30},
  year={2017},
  publisher={ACM}
}

@article{ScudderIT1965,
  title={{Probability of error of some adaptive pattern-recognition machines}},
  author={Scudder, Henry},
  journal={IEEE Transactions on Information Theory},
  volume={11},
  number={3},
  pages={363--371},
  year={1965},
  publisher={IEEE}
}

@inproceedings{ADReSS,
  title     = {{Alzheimer’s Dementia Recognition Through Spontaneous Speech: The ADReSS Challenge}},
  author    = {Saturnino Luz and Fasih Haider and Sofia de la Fuente and Davida Fromm and Brian MacWhinney},
  year      = {2020},
  booktitle = {Interspeech},
  pages     = {2172--2176},
}

@inproceedings{LuzMultilingualICASSP2023,
  title={{Multilingual alzheimer’s dementia recognition through spontaneous speech: a signal processing grand challenge}},
  author={Luz, Saturnino and Haider, Fasih and Fromm, Davida and Lazarou, Ioulietta and Kompatsiaris, Ioannis and MacWhinney, Brian},
  booktitle={International Conference on Acoustics, Speech and Signal Processing (ICASSP)},
  pages={1--2},
  year={2023},
  organization={IEEE}
}

@inproceedings{xlsr1,
  title={{Cross-lingual Alzheimer’s disease detection based on paralinguistic and pre-trained features}},
  author={Chen, Xuchu and Pu, Yu and Li, Jinpeng and Zhang, Wei-Qiang},
  booktitle={International Conference on Acoustics, Speech and Signal Processing (ICASSP)},
  pages={1--2},
  year={2023},
  organization={IEEE}
}

@inproceedings{bert,
    title = {{"BERT: Pre-training of Deep Bidirectional Transformers for Language Understanding"}},
    author = "Devlin, Jacob  and
      Chang, Ming-Wei  and
      Lee, Kenton  and
      Toutanova, Kristina",
    editor = "Burstein, Jill  and
      Doran, Christy  and
      Solorio, Thamar",
    booktitle = "Proceedings of the 2019 Conference of the North {A}merican Chapter of the Association for Computational Linguistics: Human Language Technologies, Volume 1 (Long and Short Papers)",
    month = jun,
    year = "2019",
    address = "Minneapolis, Minnesota",
    publisher = "Association for Computational Linguistics",
    pages = "4171--4186",
}

@article{Silhouette,
  title={Silhouettes: a graphical aid to the interpretation and validation of cluster analysis},
  author={Rousseeuw, Peter J},
  journal={Journal of computational and applied mathematics},
  volume={20},
  pages={53--65},
  year={1987},
  publisher={Elsevier}
}

@article{seperation,
  title={A cluster separation measure},
  author={Davies, David L and Bouldin, Donald W},
  journal={IEEE transactions on pattern analysis and machine intelligence},
  number={2},
  pages={224--227},
  year={2009},
  publisher={Ieee}
}

\end{document}